# Integration of TensorFlow based Acoustic Model with Kaldi WFST Decoder

*Minkyu Lim, Ji-Hwan Kim*

Department of Computer Science and Engineering, Sogang University, Republic of Korea
`{lmkhi, kimjihwan}@sogang.ac.kr`

## Abstract

While the Kaldi framework provides state-of-the-art components for speech recognition like feature extraction, deep neural network (DNN)-based acoustic models, and a weighted finite state transducer (WFST)-based decoder, it is difficult to implement a new flexible DNN model. By contrast, a general-purpose deep learning framework, such as TensorFlow, can easily build various types of neural network architectures using a tensor-based computation method, but it is difficult to apply them to WFST-based speech recognition. In this study, a TensorFlow-based acoustic model is integrated with a WFST-based Kaldi decoder to combine the two frameworks. The features and alignments used in Kaldi are converted so they can be trained by the TensorFlow model, and the DNN-based acoustic model is then trained. In the integrated Kaldi decoder, the posterior probabilities are calculated by querying the trained TensorFlow model, and a beam search is performed to generate the lattice. The advantages of the proposed one-pass decoder include the application of various types of neural networks to WFST-based speech recognition and WFST-based online decoding using a TensorFlow-based acoustic model. The TensorFlow based acoustic models trained using the RM, WSJ, and LibriSpeech datasets show the same level of performance as the model trained using the Kaldi framework.

## 1. Introduction

Automatic speech recognition (ASR) has been significantly improved in recent years [1, 2, 3]. Many researchers and leading companies have actively engaged in studying speech recognition. One of the reasons for this active research is that various open-source ASR frameworks have been introduced. In the early stages, there was a lot of research on using HTK and Sphinx in ASR based on Gaussian mixture model-hidden Markov models (GMM-HMMs). Recently, as deep learning has been progressively developed, deep neural network-hidden Markov models (DNN-HMMs)-based ASR studies using CNTK and Kaldi have been actively conducted [4, 5, 6, 7].

Kaldi is currently the most widely used ASR system written in C++. It uses a decoding graph based on weighted finite-state transducers (WFSTs) [8]. Kaldi helps accelerate research and development by providing recipes for various datasets throughout the process, including feature extraction, model training, decoding, and scoring. In addition to recipes, various type of neural network models such as *p-norm* DNN, long-short term memory (LSTM), time delayed neural network (TDNN) are provided [9, 10, 11]. However, it is not easy to implement a new type of neural network model that is different from the pre-defined model architectures using Kaldi, because the forward and backward propagation must be implemented based on C++ at the source code level.

While Kaldi is a deep learning framework for speech recognition, TensorFlow is one of several general-purpose deep learning frameworks, which include MXNet, PyTorch, and Theano [12, 13, 14, 15]. State-of-the-art technologies related to DNNs are applied to TensorFlow, which builds a tensor-based computation graph, and is specialized for computation through GPU parallel processing. However, it is difficult to build a WFST-based speech recognition system using TensorFlow because it requires the implementation of many components that are not supported, such as the feature extractor and decoder. Therefore, most TensorFlow-based speech recognition studies are related to convolutional neural network (CNN)-based word recognition, language modeling, and language translation [16, 17].

Recently, many studies have attempted to bridge the gap between Kaldi and general-purpose deep learning frameworks [18]. PyTorch-Kaldi is a toolkit that links PyTorch, which is a Python-based general-purpose deep learning framework, with Kaldi. In PyTorch-Kaldi, the AM is trained in PyTorch, and all other tasks such as feature extraction, labeling, and decoding are performed in Kaldi. PyTorch-Kaldi is configured so that users can easily define the DNN-, CNN-, and recurrent neural network (RNN)-related models and hyperparameters through configuration. However, in PyTorch-Kaldi, the decoding process for the input utterance is a two-pass search strategy in which the posterior probabilities of the context-dependent phone states through the PyTorch-based AM are stored separately and the stored data is loaded by the Kaldi decoder. Therefore, it is difficult to support online decoding.

In this study, a method for acoustic modeling using both the Kaldi and TensorFlow frameworks is proposed to eliminate the gap between the two. Furthermore, a method to use the output of the TensorFlow AM directly in the Kaldi decoder is proposed so that online decoding can be supported. Using this method, users can focus more on neural network architecture. The code and scripts of the proposed system are open-source.[1]

Section 2 describes the procedure for TensorFlow-based acoustic modeling. Section 3 describes a method to integrate TensorFlow-based AM with the Kaldi decoder. Section 4 describes the evaluation process and results from using the TensorFlow-based AM and integrated Kaldi decoder. Finally, Section 5 presents the conclusions.

---

Corresponding Author: Ji-Hwan Kim

[1] https://github.com/MinkyuLim/kaldi-with-tensorflow-am

## 2. TensorFlow based Acoustic Modeling

The overall procedure to build a Kaldi ASR system is shown in Figure 1. The system is divided into feature extraction, GMM training, DNN training, and WFST decoding processes. For the training step, TensorFlow is used for DNN model training and decoding Kaldi is used to obtain the feature and label information for the TensorFlow-based acoustic modeling.

The results of the feature model and label configuration obtained using Kaldi are used for the TensorFlow-based acoustic model. The format transformation is performed based on the feature extraction result from Kaldi and the alignment result, which is applied to the training dataset after the GMM training, so the TensorFlow-based model can be trained. In Figure 1, the DNN training step and the DNN model, which is learned by the training step, are integrated so that they can be used in the Kaldi decoder. A conversion process for the feature and alignment information obtained from Kaldi to the type used in TensorFlow is described is Section 2.1, and the TensorFlow-based acoustic modeling process is described in Section 2.2.

### 2.1. Features and Labels

The feature vectors and alignment results of Kaldi are stored in an *ark* format, i.e., a binary file, which is composed of key-value pairs. In general, a single utterance is assigned a single key, and the feature sequence and alignment information for the utterance is stored as the value.

Figure 2 depicts the process of the transformation of Kaldi features and labels. For each utterance, each feature vector is concatenated according to the left and right context information in each frame, and the corresponding label is extracted from the pdf index of the corresponding frame from the alignment file in Kaldi. In this module, the feature vector and alignment data stored in the Kaldi style are automatically converted by splitting the dataset according to the parameters (context length, feature dimension) provided for the user's convenience. The entire learning data is split into *N* units and dataset conversion is performed for each unit.

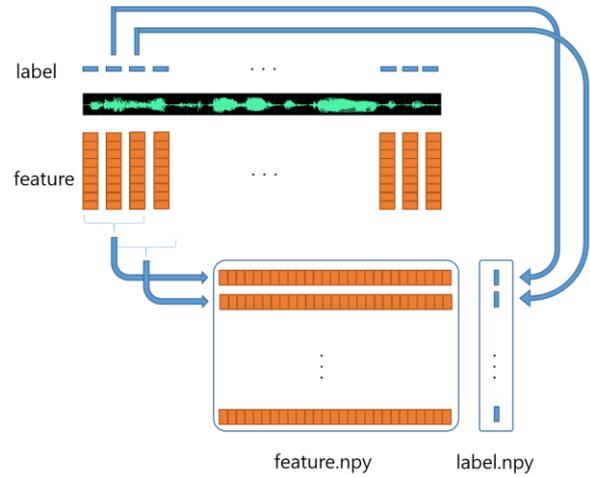

Figure 2: *Transformation of Kaldi features and labels.*

### 2.2. Acoustic Model Training with TensorFlow

Once the features and labels are prepared, TensorFlow-based acoustic model training is conducted. For the neural network model, RNN models provided in TensorFlow, such as the standard DNN, CNN, LSTM, which is a sequence model, or gated recurrent unit (GRU), are used. Furthermore, the state-of-the-art model training technics, like noise-contrastive estimation (NCE), dropout, and batch-normalization, can be used [19, 20, 21, 22]. The output of the neural network model is a probability value for the context-dependent phone state determined in the GMM learning step for the input feature vector. In this step, an archive format dataset is read, converted into a chunk unit consisting of several frames, and used in the neural net learning for the neural net training process in Kaldi and the input vector.

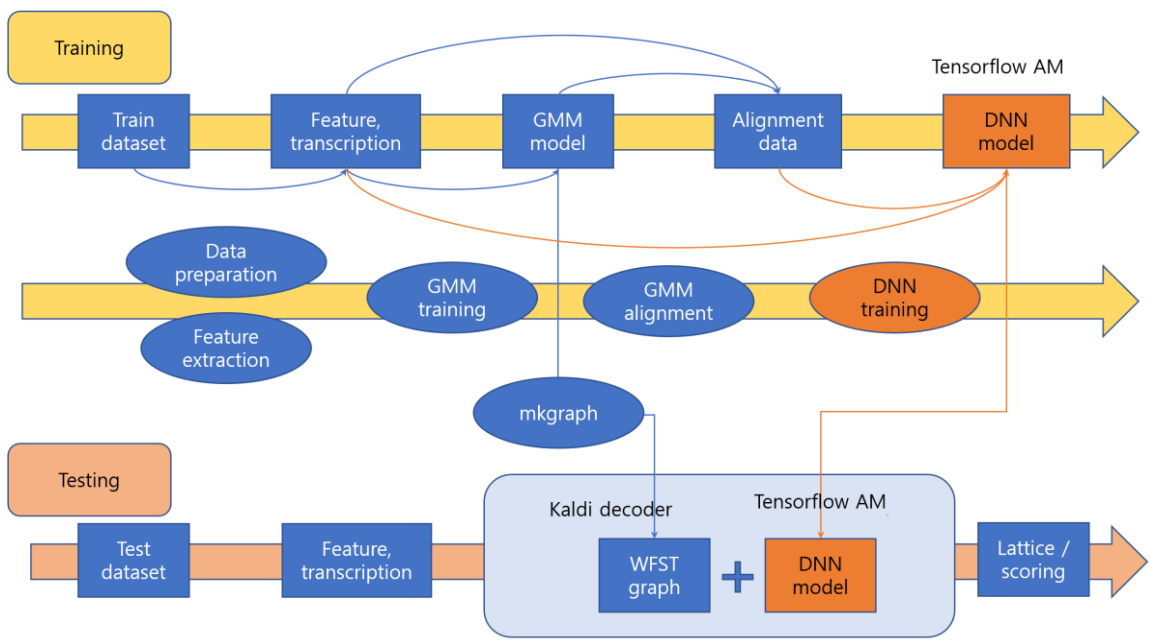

Figure 1: *Overall procedure of Kaldi speech recognition system.*

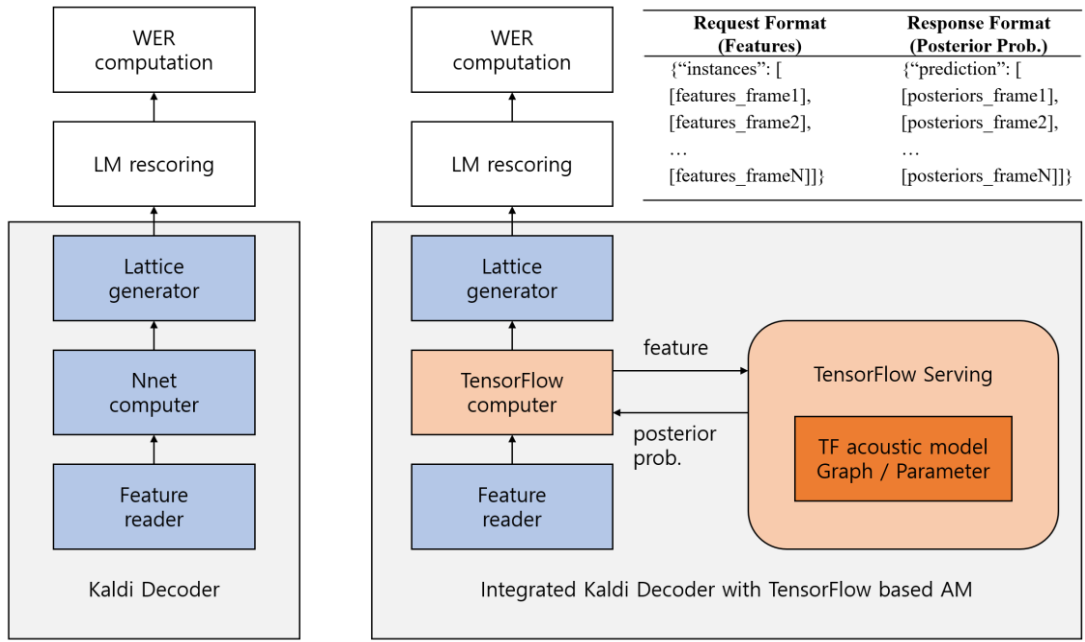

Figure 3: *Structure of Kaldi decoder and integrated Kaldi decoder with TensorFlow based AM.*

## 3. Integration with Kaldi Decoder

The Kaldi decoder consists of the *feature reader*, *nnetcomputer*, and *lattice generator* as shown in Figure 3. The *feature reader* reads a feature vector in *ark* format, reconstructs the feature vector according to the left and right context length, and transmits it to *nnetcomputer*. In *nnetcomputer*, after the posterior probabilities of context-dependent phone states for each frame are calculated through forward propagation, the prior probabilities are multiplied, which is then transmitted to the *lattice generator*. Accordingly, the *lattice generator* generates a lattice by performing the beam search algorithm. To integrate TensorFlow-based AM into Kaldi decoder, *nnetcomputer* must be replaced to a module in which the forward propagation of TensorFlow-based AM is performed.

We propose a server-client model to integrate TensorFlow-based AM into Kaldi decoder. TensorFlow Serving is used to integrate the TensorFlow-based AM into Kaldi decoder [23]. TensorFlow Serving is a toolkit that can easily deploy a trained model into a server–client architecture. When a client sends an http POST request message, which fills the input feature vector in a *json* format, the TensorFlow Serving server performs a forward propagation of the pre-loaded neural network model and returns the output value to the client.

Figure 3. shows the structure of the Kaldi decoder integrated with TensorFlow-based AM. *Nnetcomputer* is replaced with *tfcomputer*, which includes the client of TensorFlow Serving server. In *tfcomputer*, chunked feature vectors are serialized in an http POST request message in *json* format, and the message is sent to the TensorFlow Serving server. The posterior result from the server is deserialized and passed to lattice generator.

*Nnetcomputer* is a class defined in nnet-compute.cc source code. *Nnetcomputer* has two main functions: Propagate(), Backprop(). Backprop function is used only in the Kaldi train step, so it is not implemented in *tfcomputer*. In the Propagate() function of *tfcomputer*, serialization, http query, and deserialization processes are implemented. RapidJSON is used for serializing and deserializing [1], and CURL is used for querying [2], and kaldi-io-for-python is used for feature and label format transformation [3].

Table 1: *Difference between Kaldi decoder and integrated decoder with TensorFlow based AM.*

| Source Level | Kaldi Decoder | Integrated Kaldi Decoder with TensorFlow based AM |
|---|---|---|
| script | steps/nnet2/decode.sh | steps/tf/decode.sh |
| binary | nnet2bin/nnet-latgen-faster.cc | nnet2bin/tf-latgen-faster.cc |
| library | nnet-compute.cc<br>nnet-compute.h | tf-compute.cc<br>tf-compute.h |
| class | class NnetComputer {<br>  public:<br>    Propagate();<br>    Backprop();<br>  private:<br>    GetOutput();<br>} | class TfComputer {<br>  public:<br>    Propagate();<br>  private:<br>    Serialize();<br>    Deserialize();<br>    GetOutput();<br>} |

An integrated decoder with the proposed method is implemented so that it does not affect existing script, binary, and library. Therefore, users only have to call the newly implemented script to operate the TensorFlow decoder. Table 1 shows the difference between Kaldi decoder and TensorFlow decoder.

---

[1] http://rapidjson.org/
[2] https://curl.haxx.se/
[3] https://github.com/vesis84/kaldi-io-for-python

## 4. Experiments

From among the learning materials that are widely used for speech recognition evaluation, three datasets were chosen for our experiments ranging from a small corpus to a large corpus. The corpora that were used in the experiments are RM, WSJ, and LibriSpeech, whose total learning lengths are 9 hours, 80 hours, and 960 hours, respectively. In the recipe provided by Kaldi according to each dataset, the learning was conducted up to the nnet2 model learning. Feature/label set transformation was performed so that the *ark*-type features and an alignment file that were extracted from Kaldi could be used to train the TensorFlow models. Once the TensorFlow model training was completed the server was operated by using TensorFlow Serving. After this, a test was performed using the integrated Kaldi decoder, and the word error rate (WER) was measured using the Kaldi scorer.

### 4.1. RM

The baseline acoustic model that was used in the RM dataset experiment is the same model that was used in the Kaldi *nnet4d3* recipe. This model is composed of two *p-norm* layers with 1,000 neurons in each layer. The number of pdfs that correspond to the outputs of the neural network is 1,504. The data used for training the TensorFlow model was generated by using the 40-dimensional *fMLLR* features generated from the corresponding recipe and the alignment created in the *tri3b* step [24]. A multi-layer perceptron (MLP) consisting of two layers containing 1,000 neurons each was evaluated by conducting the learning with TensorFlow. Table 1 shows the WER of the two models. The TensorFlow-based acoustic model showed comparable performance to that of the Kaldi recipe.

Table 2: *WER of Kaldi baseline model and TensorFlow based acoustic model.*

| Kaldi *p-norm* AM + Kaldi Decoder | TensorFlow based MLP AM + Integrated Kaldi Decoder |
|---|---|
| 1.97 | 1.90 |

### 4.2. WSJ

The WSJ and LibriSpeech corpora were used to evaluate the TensorFlow-based MLP, RNN, and LSTM acoustic models and integrated Kaldi decoder [25, 26]. The context lengths of the input features of each model were adjusted to 15, 30, and 60 in the experiments.

Table 3: *Evaluation results of decode_bd_tgpr_eval92 for WSJ dataset.*

| Models | WER (%) Context Windows | | |
|---|---|---|---|
| | 15 | 31 | 61 |
| MLP | 4.7 | 4.6 | 4.5 |
| RNN | 4.9 | 4.8 | 4.8 |
| LSTM | 4.6 | 4.6 | 4.4 |

In the case of the WSJ corpus, the features and labels used to train the TensorFlow models were 40-dimensional *fMLLR* features that were extracted from Kaldi, and the result of Kaldi *tri4b_ali_si284* was used for the alignment. The number of pdfs was 3,352. Test dataset and decoding graph was same as the *decode_bd_tgpr_eval92* script of Kaldi recipe. The experimental results are shown in Table 2. The MLP model had four layers with 2,000 neurons each. The RNN and the LSTM models both had two layers with 500 neurons each.

### 4.3. LibriSpeech

In the Librispeech corpus, the features and labels used for training the TensorFlow models were 40-dimensional *fMLLR* features that were extracted from Kaldi, and the result of Kaldi *tri6b* was used for the alignment. The number of pdfs that were used for the neural network outputs was 5,704. Test dataset and decoding graph and rescoring LM was same as the *decode_fglarge_test_clean* script of Kaldi recipe. The experimental results are shown in Table 2. The MLP model had six layers with 5,000 neurons each. Both the RNN and LSTM models had three layers with 1,000 neurons each. This result was similar to the best result of the Kaldi *p-norm* DNN model with WER 5.5.

Table 3: *Evaluation results of decode_fglarge_test_clean for Librispeech dataset.*

| Models | WER (%) Context Windows | | |
|---|---|---|---|
| | 15 | 31 | 61 |
| MLP | 5.6 | 5.6 | 5.5 |
| RNN | 5.9 | 5.8 | 5.8 |
| LSTM | 5.6 | 5.5 | 5.4 |

## 5. Conclusion and Future Work

We proposed a method in which the gap between Kaldi and TensorFlow is eliminated, an acoustic model is trained in TensorFlow, and the model is integrated into the Kaldi-based decoder. The acoustic model training is conducted in TensorFlow, and the features, labels, and other WFST-based decoders are obtained through Kaldi. To integrate the learned model into the Kaldi decoder, TensorFlow Serving is used, and a query is sent to TensorFlow Serving to perform a neural net computation in the Kaldi decoder. The pdf probability inferred from TensorFlow Serving is transmitted to the Kaldi decoder, and a lattice is generated through its beam search. The proposed method made it possible for the various type of neural net model to learn in TensorFlow, which addressed the difficultly faced in implementing and conducting a learning process for a new type of neural net model in Kaldi. Therefore, this method is suitable for performing acoustic model studies in the future in a more flexible manner. Furthermore, as online decoding is made possible by using the TensorFlow-based acoustic model, this method can be applied to a product. In future, we plan to apply this method to Kaldi online decoding and *nnet3* / *chain* models.

## 6. Acknowledgements

This work was supported by Institute for Information & communications Technology Promotion (IITP) grant funded by the Korea government (MSIT) (No.2017-0-01772, Development of QA systems for Video Story Understanding to pass the Video Turing Test)

# 7. References

[1] Y. Lecun, Y. Bengio, and G. Hinton, "Deep learning," *nature*, vol. 521, no. 7553, pp. 436-444, 2015.

[2] J. Schmidhuber, "Deep learning in neural networks: An overview," *Neural networks*, vol. 61, pp-85-117, 2015.

[3] D. Yu and J. Li, "Recent progresses in deep learning based acoustic models," *IEEE/CAA Journal of Automatica Sinica*, vol. 4, no. 3, pp. 399–412, 2019.

[4] S. Young, G. Evermann, M. Gales, T. Hain, D. Kershaw, X. Liu, G. Moore, J. Odell, D. Ollason, D. Povey, V. Valtchev, and P. Woodland, *The HTK Book (for version 3.4)*, Cambridge University Engineering Department, 2009.

[5] P. Lamere, P. Kwok, E. Gouvea, B. Raj, R. Singh, W. Walker, M. Warmuth, and P. Wolf, "The CMU SPHINX-4 speech recognition system," in *ICAASP 2003 – 2003 IEEE International Conference on Acoustics, Speech and Signal Processing, April 6-10, Hong Kong, China, Proceedings,* 2003, pp. 2-5.

[6] F. Seide and A. Agarwal, "CNTK: Microsoft's open-source deep-learning toolkit," in *SIGKDD 2016 – 22nd ACM SIGKDD Conference on Knowledge Discovery and Data Mining, August 13-17, San Francisco, U.S.A., Proceedings,* 2016, pp. 2135-2135

[7] D. Povey, A. Ghoshal, G. Boulianne, L. Burget, O. Glembek, N. Goel, M. Hannemann, P. Motlicek, Y. Qian, P. Schwarz, J. Silovsky, G. Stemmer, and K. Vesely, "The Kaldi Speech Recognition Toolkit," *2011 IEEE Workshop on Automatic Speech Recognition and Understanding,* IEEE Signal Processing Society, 2011.

[8] M. Mohri, F. Pereira, and M. Riley, "Weighted finite-state transducers in speech recognition," *Computer Speech and Language*, vol. 16, no. 1, pp. 69-88, 2002.

[9] X. Zhang, J. Trmal, D. Povey, and S. Khudanpur, "Improving deep neural network acoustic models using generalized maxout networks," in *ICASSP 2014 - 2014 IEEE International Conference on Acoustics, Speech and Signal Processing, May 4-9, Florence, Italy, Proceedings,* 2014, pp. 215-219.

[10] G. Cheng, V. Peddinti, D. Povey, V. Manohar, S. Khudanpur, and Y. Yan, "An Exploration of Dropout with LSTMs," in *INTERSPEECH 2017 – 18th Annual Conference of the International Speech Communication Association, August 20-24, Stockholm, Sweden, Proceedings*, 2017, pp. 1586-1590.

[11] V. Peddinti, D. Povey, and S. Khudanpur, "A time delay neural network architecture for efficient modeling of long temporal contexts," in *INTERSPEECH 2015 – 16th Annual Conference of the International Speech Communication Association, September 6-10, Dresden, Germany, Proceedings,* 2015, pp. 3214-3218.

[12] M. Abadi, P. Barham, J. Chen, Z. Chen, A. Davis, J. Dean, and M. Kudlur, "Tensorflow: A system for large-scale machine learning," in *OSDI 2016 - 12th USENIX Symposium on Operating Systems Design and Implementation, November 2-4, Savannah, U.S.A.,* 2016, pp. 265-283.

[13] R. Collobert, S. Bengio, and J. Mariéthoz, "Torch: a modular machine learning software library," *Technical Report IDIAP-RR 02-46*, 2002.

[14] N. Ketkar, *"Introduction to pytorch." In Deep learning with python*, Apress, 2017.

[15] J. Bergstra, O. Breuleux, F. Bastien, P. Lamblin, R. Pascanu, G. Desjardins J. Turian, D. Farley, and Y. Bengio, "Theano: a CPU and GPU math expression compiler," in *SciPy 2010 - Proceedings of the Python for Scientific Computing Conference, June 28-July 3, Austin, U.S.A., Proceedings,* 2010, pp 3-10.

[16] J. Chorowski and N. Jaitly, "Towards better decoding and language model integration in sequence to sequence models," *arXiv preprint arXiv:1612.02695*, 2016.

[17] Y. Wu, M. Schuster, Z. Chen, V. Le, M. Norouzi, W. Macherey, and J. Klingner, "Google's neural machine translation system: Bridging the gap between human and machine translation," *arXiv preprint arXiv:1609.08144*, 2016.

[18] M. Ravanelli, T. Parcollet, and Y. Bengio, "The PyTorch-Kaldi Speech Recognition Toolkit," *arXiv preprint arXiv:1811.07453*, 2018.

[19] M. Gutmann and A. Hyvärinen "Noise-contrastive estimation: A new estimation principle for unnormalized statistical models," in *AISTATS 2010 - 13th International Conference on Artificial Intelligence and Statistics, May 13-15, Sardinia, Italy, Proceedings,* 2010, pp. 297-304.

[20] N. Srivastava, G. Hinton, A. Krizhevsky, I. Sutskever, and R. Salakhutdinov, "Dropout: a simple way to prevent neural networks from overfitting," *The Journal of Machine Learning Research*, vol. 15, no. 1, pp. 1929-1958, 2014.

[21] S. Ioffe and C. Szegedy, "Batch normalization: Accelerating deep network training by reducing internal covariate shift," *arXiv preprint arXiv:1502.03167*, 2015.

[22] S. Ruder, "An overview of gradient descent optimization algorithms," *arXiv preprint arXiv:1609.04747*, 2016.

[23] C. Olston, N. Fiedel, K. Gorovoy, J. Harmsen, L. Lao, F. Li, and J. Soyke, "Tensorflow-serving: Flexible, high-performance ml serving," *arXiv preprint arXiv:1712.06139*, 2017.

[24] D. Povey, D. Kanevsky, B. Kingsbury, B. Ramabhadran, G. Saon, and K. Visweswariah, "Boosted MMI for model and feature-space discriminative training," in *ICASSP 2008 – 2008 IEEE International Conference on Acoustics, Speech and Signal Processing, March 30-April 4, Las Vegas, U.S.A., Proceedings,* 2008, pp. 4057-4060.

[25] D. Paul and J. Baker, "The design for the Wall Street Journal-based CSR corpus," in *Proceedings of the workshop on Speech and Natural Language, February 23-26, Harriman, U.S.A., Proceedings,* 1992.

[26] V. Panayotov, G. Chen, D. Povey, and S. Khudanpur, "Librispeech: an ASR corpus based on public domain audio books," in *ICASSP 2015 – 2015 IEEE International Conference on Acoustics, Speech and Signal Processing, April 19-24, Brisbane, Australia, Proceedings,* 2015, pp. 5206-5210.